# Self-Equalization of Energies of Solitons in Transmission Systems by Guiding Filters

S. M. Shahruz

Berkeley Engineering Research Institute
P. O. Box 9984
Berkeley, California 94709

## Abstract

In this paper, the regulating effect of guiding filters on the energies of solitons in wave-length division multiplexing (WDM) transmission systems is rigorously studied. More precisely, it is shown that guiding filters prevent the energies of solitons from decaying to zero in long distances. This goal is achieved by studying a mathematical model of the evolution of energies of solitons in transmission systems. The approach is mathematical and does not use numerical or experimental results available in the literature; nevertheless, it reaffirms such results.

**Keywords:** Soliton wave-length division multiplexing (WDM) transmission systems, Guiding filters, Soliton energies, Nonlinear dynamical systems, Lyapunov's linearization method.

## 1. Introduction

In multi-channel wave-length division multiplexing (WDM) transmission systems, due to the variation of amplifier gain with the wavelength, it is impossible to maintain the energies of transmitted signals in different channels almost equal. The differences of energies of signals can grow significantly large when the transmission length increases. For instance, the energies of signals in some channels tend to zero in long distances.

The problem of different energies of signals and decaying of some of them to zero in long distances can be overcome by guiding filters. Guiding filters are commonly used to suppress

Gordon-Haus jitter and other noise effects in soliton transmission systems; see, e.g., Refs. [1-8] and references therein. Such filters can also counter the effects of variable amplifier gain and regulate the energies of solitons in channels of a soliton transmission system. It has been demonstrated by numerical simulation and observed experimentally that guiding filters provide a feedback mechanism that locks the energies of solitons in different channels to fixed values that do not change with distance, even in the presence of large amplifier gain variations; see Refs. [5, 8]. Roughly speaking, guiding filters provide a loss that is proportional to the soliton energy. Therefore, if the soliton energy increases due to an increase in the channel amplifier gain, then the loss increases in order to limit the soliton energy growth. On the other hand, for a channel with small gain, the loss drops, and hence the soliton energy does not decay. The regulating effect of guiding filters has played a key role in successful error free soliton WDM transmission over transoceanic distances; see, e.g., Refs. [6-8].

In this paper, the regulating effect of guiding filters in preventing the energies of solitons from decaying to zero in long distances is studied rigorously. In other words, results from numerical studies and experimental observations by researchers that show the success of guiding filters are all proved mathematically in this paper. The organization of the paper is as follows. In Section 2, models of the evolution of energies of solitons in a soliton WDM transmission system are presented. In Section 3, transmission of solitons in the absence of guiding filters is studied. It is shown that the energies of solitons in all channels, except in one, tend to zero in long distances. In Section 4, it is shown that when guiding filters are used the energies of solitons in all channels remain positive in long distances.

## 2. Mathematical Models

In this section, mathematical models of the evolution of energies of solitons in a soliton WDM transmission system with $n$ channels are presented. These models have been studied by numerical simulation and experimentally in Refs. [5, 8].

Let $W_i$ denote the soliton pulse energy in the $i$-th channel for an $i = 1, 2, \ldots, n$ when guiding filters are used. The energy loss of the soliton due to the guiding filter in the $i$-th channel is a monotonically increasing function of $W_i/D_i$, where $D_i$ denotes the path-average dispersion; see, e.g., Ref. [7]. This function is denoted by $f_i$. For Gaussian filters, $f_i$ is a quadratic function of $W_i/D_i$. For shallow etalon filters, which are widely used in practice, $f_i$ is a linear function given by

$$f_i\left(\frac{W_i}{D_i}\right) = \frac{s_i W_i}{D_i}, \tag{2.1}$$

where $s_i > 0$ is a constant real number for all $i = 1, 2, \ldots, n$. The evolution of $W_1, W_2, \ldots, W_n$ with respect to distance $z$ can be represented by the following coupled nonlinear ordinary differential equations (Refs. [5] and [8, pp. 142-144]):

$$\frac{dW_1}{dz} = \frac{g_1 W_1}{1 + C(W_1 + W_2 + \ldots + W_n)} - l_1 W_1 - \frac{s_1 W_1^2}{D_1}, \quad W_1(0) = W_{10} > 0, \tag{2.2.1}$$

$$\frac{dW_2}{dz} = \frac{g_2 W_2}{1 + C(W_1 + W_2 + \ldots + W_n)} - l_2 W_2 - \frac{s_2 W_2^2}{D_2}, \quad W_2(0) = W_{20} > 0, \tag{2.2.2}$$

$$\vdots$$

$$\frac{dW_n}{dz} = \frac{g_n W_n}{1 + C(W_1 + W_2 + \ldots + W_n)} - l_n W_n - \frac{s_n W_n^2}{D_n}, \quad W_n(0) = W_{n0} > 0, \tag{2.2.n}$$

for all $z \in [0, z_f]$, where $z_f$ is the length of the transmission system. In (2.2), parameters $g_i > 0$ and $l_i > 0$ are constant real numbers for all $i = 1, 2, \ldots, n$, which denote, respectively, the small-signal gain rate and the linear loss rate in the $i$-th channel, and

$$C = \frac{m R}{P_{sat}}, \tag{2.3}$$

is a constant positive real number, where $m > 0$ is the mark-to-space ratio (usually $0.5$), $R > 0$ is the per-channel bit rate, and $P_{sat} > 0$ is the saturation power of the amplifier. In (2.2), the positive real number $W_{i0}$ denotes the initial condition of $W_i$ for an $i = 1, 2, \ldots, n$. The last terms on the right-hand sides of (2.2) are due to guiding filters.

In the absence of guiding filters, a set of equations other than (2.2) describes the evolution of energies of solitons. Let $V_i$ denote the soliton pulse energy in the $i$-th channel for an $i = 1, 2, \ldots, n$ when guiding filters are not used. The evolution of $V_1, V_2, \ldots, V_n$ with respect to distance $z$ can be represented by the following coupled nonlinear ordinary differential equations:

$$\frac{dV_1}{dz} = \frac{g_1 V_1}{1 + C (V_1 + V_2 + \ldots + V_n)} - l_1 V_1, \quad V_1(0) = V_{10} > 0, \quad (2.4.1)$$

$$\frac{dV_2}{dz} = \frac{g_2 V_2}{1 + C (V_1 + V_2 + \ldots + V_n)} - l_2 V_2, \quad V_2(0) = V_{20} > 0, \quad (2.4.2)$$

$$\vdots$$

$$\frac{dV_n}{dz} = \frac{g_n V_n}{1 + C (V_1 + V_2 + \ldots + V_n)} - l_n V_n, \quad V_n(0) = V_{n0} > 0, \quad (2.4.n)$$

where parameters $g_i$ and $l_i$ are the same as those in (2.2) for all $i = 1, 2, \ldots, n$, the constant number $C$ is the same as that in (2.3), and the positive real number $V_{i0}$ denotes the initial condition of $V_i$ for an $i = 1, 2, \ldots, n$.

Both systems (2.2) and (2.4) share the following important property, which is obvious from the physical point of view, but can be established mathematically.

**Theorem 2.1:** The soliton pulse energy $W_i(z)$ (respectively, $V_i(z)$) in system (2.2) ((2.4)) is non-negative for all $z \in [0, z_f]$ and $i = 1, 2, \ldots, n$.

**Proof:** The right-hand side of (2.2.i) (respectively, (2.4.i)) is zero at $W_i = 0$ ($V_i = 0$) for all $i = 1, 2, \ldots, n$. Thus, by Proposition 1.1 in Ref. [9, p. 21], the proof follows. □

The steady-state values of $W_i(z)$ and $V_i(z)$ for all $i = 1, 2, \ldots, n$ are of great interest as the transmission length increases. Therefore, it is assumed that the transmission length is infinitely large; that is, $z_f \to \infty$. With this assumption and Theorem 2.1, systems (2.2) and (2.4) can be regarded as dynamical systems whose trajectories are in the positive orthant $\mathbb{R}_+^n$. The steady-state values, or the equilibrium points of systems (2.2) and (2.4), are points in $\mathbb{R}_+^n$, denoted, respectively, by

$$W^* := (W_1^*, W_2^*, \ldots, W_n^*) \in \mathbb{R}_+^n, \quad (2.5a)$$

$$V^* := (V_1^*, V_2^*, \ldots, V_n^*) \in \mathbb{R}_+^n, \quad (2.5b)$$

which render the right-hand sides of (2.2) and (2.4) zero.

A definition to be used in the following sections is now given.

**Definition 2.2:** An equilibrium point of system (2.2) or system (2.4) is called a type-$k$ equilibrium point for an integer $1 \leq k \leq n$, when only $k$ of its coordinates are positive (and the rest of them are zero). □

Up to this point, two nonlinear dynamical systems (2.4) and (2.2) were presented that, respectively, describe the evolution of energies of solitons in transmission systems in the absence and presence of guiding filters. The equilibrium points of these systems represent the energies of solitons in long distances. Therefore, it is important to know which equilibrium points can be attained by systems (2.4) and (2.2). In the next two sections, these points are carefully studied.

## 3. Transmission in the Absence of Guiding Filters

In this section, transmission of solitons in an $n$-channel system in the absence of guiding filters is studied via (2.4). More precisely, the soliton pulse energies in channels in long distances are studied. As it was stated earlier, such energies are the equilibrium points of system (2.4). These points are first obtained and then it is determined whether they are stable. In the following, $\theta_n$ denotes the origin in $\mathbb{R}_+^n$, which is a point whose coordinates are all zero.

### 3.1. Equilibrium Points

It is first noted that the origin is an equilibrium point of system (2.4).

**Fact 3.1:** The origin $\theta_n \in \mathbb{R}_+^n$ is an equilibrium point of system (2.4).

**Proof:** Obvious. □

Next, it is shown that system (2.4) has at most $n$ type-1 equilibrium points.

**Fact 3.2:** If for any $i = 1, 2, \ldots, n$, the ratio

$$\frac{g_i}{l_i} > 1, \tag{3.1}$$

then system (2.4) has a type-1 equilibrium point at

$$E_i^* = \left(0, \ldots, 0, \frac{1}{C}(\frac{g_i}{l_i} - 1), 0, \ldots, 0\right) \in \mathbb{R}_+^n, \tag{3.2}$$

where the $i$-th coordinate is positive.

**Proof:** The point $E_i^*$ in (3.2) renders the right-hand sides of (2.4) zero. Thus, $E_i^*$ is an equilibrium point, where by (3.1), its $i$-th component is positive. □

**Remark:** Condition (3.1) implies that the gain rate is larger than the loss rate in the $i$-th channel. This condition can hold for any number of channels. That is, there can be as many type-1 equilibrium points as there are channels for which (3.1) holds - at most $n$ such points. □

System (2.4) can possibly have type-$k$ equilibrium points for any $2 \leq k \leq n$; they are given in the following.

**Fact 3.3:** For system (2.4), without loss of generality, assume that

$$\frac{g_1}{l_1} = \frac{g_2}{l_2} = \ldots = \frac{g_k}{l_k} > 1, \tag{3.3}$$

where $2 \leq k \leq n$. System (2.4) has type-$k$ equilibrium points at

$$V^* = (V_1^*, V_2^*, \ldots, V_k^*, 0, \ldots, 0) \in \mathbb{R}_+^n, \tag{3.4}$$

where $V_1^*, V_2^*, \ldots, V_k^*$ are any positive real numbers satisfying

$$V_1^* + V_2^* + \ldots + V_k^* = \frac{1}{C}\left(\frac{g_1}{l_1} - 1\right) = \frac{1}{C}\left(\frac{g_2}{l_2} - 1\right) = \ldots = \frac{1}{C}\left(\frac{g_k}{l_k} - 1\right) > 0. \tag{3.5}$$

**Proof:** The points $V^*$ in (3.4), whose coordinates satisfy (3.5), render the right-hand sides of (2.4) zero. Thus, they are equilibrium points. □

**Remarks: 1)** If (3.3) holds for $k = n$, then the equilibrium points in (3.4) have all their coordinates positive. It is desirable to attain such an equilibrium point since positive coordinates imply positive steady-state soliton energies in all channels.

**2)** Condition (3.3) requires that the ratios of the gain rates to the loss rates to be the same for some (or all) channels. In practice, (3.3) does not hold. Thus, type-$k$ equilibrium points for all $2 \leq k \leq n$ do not exist. If (3.3) did hold, then there would have been uncountably many (non-isolated) equilibrium points □

Up to this point, the equilibrium points of system (2.4) were obtained. Next, the stability of these points is studied.

### 3.2. Stability of Equilibrium Points

In studying the stability of equilibrium points of systems, the following technique is commonly used. The local stability and instability of an equilibrium point of a system is decided upon by the eigenvalues of the Jacobian matrix of the system at that equilibrium point. If all eigenvalues of the Jacobian matrix have negative real parts, then the equilibrium point is locally asymptotically stable. If at least one eigenvalue has positive real part, then the equilibrium point is unstable. If at least one eigenvalue of the Jacobian matrix at an equilibrium point of the system has zero real part, then the stability of that point cannot be decided upon. This technique is based on Lyapunov's linearization method; see, e.g., Ref. [9, pp. 213-214] or Ref. [10, Chap. 4].

Using the Jacobian matrix, it is first shown that $\theta_n$ cannot be attained in practice.

**Theorem 3.4:** If $g_i > l_i$ for an $i = 1, 2, \ldots, n$ (respectively, $g_i < l_i$ for all $i = 1, 2, \ldots, n$), then the origin $\theta_n$ is an unstable (a locally asymptotically stable) equilibrium point of system (2.4).

**Proof:** The Jacobian matrix of system (2.4) at $\theta_n$ is obtained as:

$$J(\theta_n) = diag\,[g_1 - l_1\,,\, g_2 - l_2\,,\, \ldots,\, g_n - l_n] \in \mathbb{R}^{n \times n}. \qquad (3.6)$$

From (3.6), the proof follows. $\square$

**Remark:** Since in practice, for at least one channel the gain rate is larger than the loss rate, by Theorem 3.4, the soliton pulse energies in all channels do not tend to zero. $\square$

The stability of type-1 equilibrium points in (3.2) is studied next.

**Theorem 3.5:** Consider system (2.4). For an integer $1 \leq i^* \leq n$, let

$$\frac{g_{i^*}}{l_{i^*}} > 1. \qquad (3.7)$$

Furthermore, let $g_{i^*}/l_{i^*}$ be the largest ratio, that is,

$$\frac{g_{i^*}}{l_{i^*}} > \frac{g_i}{l_i}, \qquad (3.8)$$

for all $i \in \{1, 2, \ldots, n\} \setminus i^*$. The type-1 equilibrium point

$$E_{i^*}^* = \left(0, \ldots, 0, \frac{1}{C}\left(\frac{g_{i^*}}{l_{i^*}} - 1\right), 0, \ldots, 0\right) \in \mathbb{R}_+^n, \qquad (3.9)$$

whose $i^*$-th coordinate is positive, is locally asymptotically stable, and all other type-1 equilibrium points are unstable.

**Proof:** The Jacobian matrix of system (2.4) at $E_{i^*}^*$ is obtained as:

$$J(E_{i^*}^*) = \begin{bmatrix} J_{11} & J_{12} \\ \Theta & J_{22} \end{bmatrix} \in \mathbb{R}^{n \times n}, \qquad (3.10)$$

where $\Theta$ is the $(n - i^*) \times i^*$ zero matrix and

$$J_{11} = \begin{bmatrix} d_1 & 0 & \cdots & 0 & 0 \\ 0 & d_2 & \cdots & 0 & 0 \\ \vdots & \vdots & \vdots & \vdots & \vdots \\ 0 & 0 & \cdots & d_{i^*-1} & 0 \\ a_{i^*} & a_{i^*} & \cdots & a_{i^*} & a_{i^*} \end{bmatrix} \in \mathbb{R}^{i^* \times i^*}, \quad J_{12} = \begin{bmatrix} 0 & 0 & \cdots & 0 \\ 0 & 0 & \cdots & 0 \\ \vdots & \vdots & \vdots & \vdots \\ 0 & 0 & \cdots & 0 \\ a_{i^*} & a_{i^*} & \cdots & a_{i^*} \end{bmatrix} \in \mathbb{R}^{i^* \times (n-i^*)}, \quad (3.11\text{a})$$

$$J_{22} = \text{diag}[d_{i^*+1}, \ldots, d_n] \in \mathbb{R}^{(n-i^*) \times (n-i^*)}, \qquad (3.11\text{b})$$

with

$$a_{i^*} = -l_{i^*}\left(1 - \frac{l_{i^*}}{g_{i^*}}\right), \qquad (3.12)$$

and

$$d_i = g_i \left(\frac{l_{i^*}}{g_{i^*}} - \frac{l_i}{g_i}\right), \qquad (3.13)$$

for all $i \in \{1, 2, \ldots, n\} \setminus i^*$.

Clearly, the eigenvalues of $J(E_{i^*}^*)$ are $a_{i^*}$ and $d_i$ for all $i \in \{1, 2, \ldots, n\} \setminus i^*$. By (3.7) and (3.8), these eigenvalues are negative. Hence, $E_{i^*}^*$ is locally asymptotically stable.

For an integer $1 \leq j \leq n$, where $j \neq i^*$, let $g_j/l_j > 1$. By Fact 3.2, the following point is also an equilibrium point of system (2.4):

$$E_j^* = \left(0, \ldots, 0, \frac{1}{C}\left(\frac{g_j}{l_j} - 1\right), 0, \ldots, 0\right) \in \mathbb{R}_+^n, \qquad (3.14)$$

where the $j$-th coordinate is positive.

The Jacobian matrix of system (2.4) at $E_j^*$ is obtained as:

$$J(E_j^*) = \begin{bmatrix} K_{11} & K_{12} \\ \Theta & K_{22} \end{bmatrix} \in \mathbb{R}^{n \times n}, \quad (3.15)$$

where $\Theta$ is the $(n-j) \times j$ zero matrix and

$$K_{11} = \begin{bmatrix} \delta_1 & 0 & \cdots & 0 & 0 \\ 0 & \delta_2 & \cdots & 0 & 0 \\ \vdots & \vdots & \vdots & \vdots & \vdots \\ 0 & 0 & \cdots & \delta_{j-1} & 0 \\ \alpha_j & \alpha_j & \cdots & \alpha_j & \alpha_j \end{bmatrix} \in \mathbb{R}^{j \times j}, \quad K_{12} = \begin{bmatrix} 0 & 0 & \cdots & 0 \\ 0 & 0 & \cdots & 0 \\ \vdots & \vdots & \vdots & \vdots \\ 0 & 0 & \cdots & 0 \\ \alpha_j & \alpha_j & \cdots & \alpha_j \end{bmatrix} \in \mathbb{R}^{j \times (n-j)}, \quad (3.16a)$$

$$K_{22} = diag[\delta_{j+1}, \ldots, \delta_n] \in \mathbb{R}^{(n-j) \times (n-j)}, \quad (3.16b)$$

with

$$\alpha_j = -l_j \left( 1 - \frac{l_j}{g_j} \right), \quad (3.17)$$

and

$$\delta_i = g_i \left( \frac{l_j}{g_j} - \frac{l_i}{g_i} \right), \quad (3.18)$$

for all $i \in \{1, 2, \ldots, n\} \setminus j$.

The eigenvalues of $J(E_j^*)$ are $\alpha_j$ and $\delta_i$ for all $i \in \{1, 2, \ldots, n\} \setminus j$. Since by (3.8), $g_{i^*}/l_{i^*} > g_j/l_j$, the eigenvalue $\delta_{i^*}$ is positive. Hence, $E_j^*$ is unstable. $\square$

**Remark:** Theorem 3.5 provides a simple procedure by which the evolution of energies of solitons in different channels in the absence of guiding filters is determined. This procedure consists of straightforward computation of the ratios of the gain rates to the loss rates for all channels. The channel, for which this ratio is larger than one and is the largest as well, achieves a positive steady-state soliton energy while the energies of solitons in all other channels tend to zero (survival of the strongest).

In Ref. [5, 8], the evolution of energies of solitons in channels is studied numerically by simulating system (2.4) when $n = 3$. The simulation results, when guiding filters are not used, are depicted in Fig. 2 of Ref. [5] or Fig. 5.13 of Ref. [8, p. 144]. From these figures, it is evident that the energies of solitons in two channels tend to zero, while the soliton energy in one channel increases and settles at a positive value. Such a behavior is certainly predicted and justified by Theorem 3.5. □

Finally, the stability of equilibrium points in (3.4) is studied.

**Theorem 3.6:** Consider system (2.4) and let (3.3) hold. The stability of any type-$k$ equilibrium point of system (2.4) given in (3.4) for all $2 \leq k \leq n$ cannot be decided upon by the eigenvalues of the Jacobian matrix of system (2.4) at that equilibrium point.

**Proof:** The Jacobian matrix of system (2.4) at $V^*$ in (3.4) is obtained as:

$$J(V^*) = \begin{bmatrix} L_{11} & L_{12} \\ \Theta & L_{22} \end{bmatrix} \in \mathbb{R}^{n \times n}, \qquad (3.19)$$

where $\Theta$ is the $(n-k) \times k$ zero matrix and

$$L_{11} = C \begin{bmatrix} -\dfrac{l_1^2 V_1^*}{g_1} & -\dfrac{l_1^2 V_1^*}{g_1} & \cdots & -\dfrac{l_1^2 V_1^*}{g_1} \\ -\dfrac{l_2^2 V_2^*}{g_2} & -\dfrac{l_2^2 V_2^*}{g_2} & \cdots & -\dfrac{l_2^2 V_2^*}{g_2} \\ \vdots & \vdots & \vdots & \vdots \\ -\dfrac{l_k^2 V_k^*}{g_k} & -\dfrac{l_k^2 V_k^*}{g_k} & \cdots & -\dfrac{l_k^2 V_k^*}{g_k} \end{bmatrix} \in \mathbb{R}^{k \times k}, \qquad (3.20a)$$

$$L_{12} = C \begin{bmatrix} -\dfrac{l_1^2 V_1^*}{g_1} & -\dfrac{l_1^2 V_1^*}{g_1} & \cdots & -\dfrac{l_1^2 V_1^*}{g_1} \\ -\dfrac{l_2^2 V_2^*}{g_2} & -\dfrac{l_2^2 V_2^*}{g_2} & \cdots & -\dfrac{l_2^2 V_2^*}{g_2} \\ \vdots & \vdots & \vdots & \vdots \\ -\dfrac{l_k^2 V_k^*}{g_k} & -\dfrac{l_k^2 V_k^*}{g_k} & \cdots & -\dfrac{l_k^2 V_k^*}{g_k} \end{bmatrix} \in \mathbb{R}^{k \times (n-k)}, \qquad (3.20b)$$

$$L_{22} = diag\left[ \dfrac{g_{k+1}}{g_1/l_1} - l_{k+1}, \dfrac{g_{k+2}}{g_1/l_1} - l_{k+2}, \ldots, \dfrac{g_n}{g_1/l_1} - l_n \right] \in \mathbb{R}^{(n-k) \times (n-k)}. \qquad (3.20c)$$

The first $k$ columns of $J(V^*)$ are identical. Thus, the rank of $J(V^*)$ is $n-k+1$ and $k-1$ of its eigenvalues are zero. Hence, the local stability of the equilibrium point (3.4) cannot be decided upon. □

**Remark:** Theorem 3.6 establishes the fact that the local stability of type-$k$ equilibrium points of system (2.4) for all $2 \leq k \leq n$ cannot be decided upon easily. There is, however, no need to try other techniques hoping to possibly establish the stability of such points. As it was reasoned earlier, due to condition (3.3), type-$k$ equilibrium points for all $2 \leq k \leq n$ do not exist in practice. □

Up to this point, system (2.4) was studied. This system represents the evolution of soliton pulse energies in an $n$-channels WDM transmission system when guiding filters are not used. It was shown that system (2.4) has only one stable type-1 equilibrium point. All coordinates of this point are zero, except one which is positive. Therefore, in the WDM transmission system, the energies of solitons in all channels, except in one, tend to zero in long distances. This is certainly an undesirable situation. The desirable situation is when system (2.4) has at least one stable type-$n$ equilibrium point. In this case, the energies of solitons in all channels remain positive even in long distances. However, type-$n$ equilibrium points do not exist for system (2.4), because the unrealistic condition (3.3) does not hold in practice.

Here is where guiding filters can enter as remedy. It is shown in the next section that guiding filters can indeed keep the energies of solitons positive in all channels of a WDM transmission system even in long distances.

## 4. Transmission in the Presence of Guiding Filters

The success of guiding filters in transmitting solitons while keeping their energy positive in long distances is shown in this section. When guiding filters are used, the soliton pulse energies in channels of a WDM transmission system in long distances are the equilibrium points of system (2.2). In this section, these points are first obtained and then their stability is studied.

### 4.1. Equilibrium Points

It is first noted that the origin is an equilibrium point of system (2.2).

**Fact 4.1:** The origin $\theta_n \in \mathbb{R}_+^n$ is an equilibrium point of the system (2.2).

**Proof:** Obvious. □

System (2.2) can have many type-$k$ equilibrium points for an integer $1 \leq k \leq n$. A type-$k$ equilibrium point of system (2.2) in $\mathbb{R}_+^n$ has $k$ positive coordinates $W_{i_1}^*, W_{i_2}^*, \ldots, W_{i_k}^*$, where each subscript $i_1, i_2, \ldots, i_k$ assumes all values $1, 2, \ldots, n$, while

$$i_1 < i_2 < \ldots < i_k . \tag{4.1}$$

Any possible sets of $i_1, i_2, \ldots, i_k$ is denoted by

$$I = \{ i_1, i_2, \ldots, i_k \} . \tag{4.2}$$

An example is now given to illustrate how many non-trivial equilibrium points are possible for system (2.2).

**Example 4.2:** For $n = 3$, all possible non-trivial equilibrium points of system (2.2) and sets $I$ are

$$W^* = (W_1^*, 0, 0), \quad I = \{ 1 \}, \tag{4.3.1}$$

$$W^* = (0, W_2^*, 0), \quad I = \{ 2 \}, \tag{4.3.2}$$

$$W^* = (0, 0, W_3^*), \quad I = \{ 3 \}, \tag{4.3.3}$$

$$W^* = (W_1^*, W_2^*, 0), \quad I = \{ 1, 2 \}, \tag{4.3.4}$$

$$W^* = (W_1^*, 0, W_3^*), \quad I = \{ 1, 3 \}, \tag{4.3.5}$$

$$W^* = (0, W_2^*, W_3^*), \quad I = \{ 2, 3 \}, \tag{4.3.6}$$

$$W^* = (W_1^*, W_2^*, W_3^*), \quad I = \{ 1, 2, 3 \} . \tag{4.3.7}$$

Clearly, there is possibly a unique type-3 equilibrium point. □

An assumption is now made. This assumption holds in practice.

**Assumption 4.3:** It is assumed that

$$\frac{g_i}{\Delta} - l_i > 0 , \tag{4.4}$$

for all $i \in I$, where $I$ is every possible set given by (4.2), and

$$\Delta = \frac{1}{2} \left( \left( 1 - C \sum_{i \in I} \frac{D_i l_i}{s_i} \right) + \left[ \left( 1 - C \sum_{i \in I} \frac{D_i l_i}{s_i} \right)^2 + 4C \sum_{i \in I} \frac{D_i g_i}{s_i} \right]^{1/2} \right) . \tag{4.5}$$

□

The positive coordinates of type-$k$ equilibrium points are now determined.

**Fact 4.4:** Consider a set $I$ given by (4.2). If

$$\sum_{i \in I} \frac{D_i(l_i - g_i)}{s_i} < 0, \tag{4.6}$$

then the positive coordinate of type-$k$ equilibrium points for an $i \in I$ is given by

$$W_i^* = \frac{D_i}{s_i}\left(\frac{g_i}{\Delta} - l_i\right), \tag{4.7}$$

where $\Delta$ is that defined in (4.5).

**Proof:** From (2.2), it follows that at an equilibrium point

$$\frac{g_i W_i^*}{1 + C \sum_{i \in I} W_i^*} - l_i^* W_i^* - \frac{s_i W_i^{*2}}{D_i} = 0, \tag{4.8}$$

for all $i \in I$. From (4.8), it is follows that

$$\frac{D_i g_i / s_i}{1 + C \sum_{i \in I} W_i^*} = \frac{D_i l_i}{s_i} + W_i^*, \tag{4.9}$$

for all $i \in I$. Summing the left- and right-hand sides of (4.9) for all $i \in I$, it is concluded that

$$\frac{\sum_{i \in I} D_i g_i / s_i}{1 + C \sum_{i \in I} W_i^*} = \sum_{i \in I} \frac{D_i l_i}{s_i} + \sum_{i \in I} W_i^*. \tag{4.10}$$

From (4.10), the following quadratic equation is obtained:

$$C\left(\sum_{i \in I} W_i^*\right)^2 + \left(1 + C \sum_{i \in I} \frac{D_i l_i}{s_i}\right)\left(\sum_{i \in I} W_i^*\right) + \sum_{i \in I} \frac{D_i(l_i - g_i)}{s_i} = 0. \tag{4.11}$$

By inequality (4.6), (4.11) has a unique positive solution for $\sum_{i \in I} W_i^*$. This solution is given by

$$\sum_{i \in I} W_i^* = \frac{1}{2C}\left(-\left(1 + C \sum_{i \in I} \frac{D_i l_i}{s_i}\right) + \left[\left(1 - C \sum_{i \in I} \frac{D_i l_i}{s_i}\right)^2 + 4C \sum_{i \in I} \frac{D_i g_i}{s_i}\right]^{1/2}\right).$$

(4.12)

Substituting $\sum_{i \in I} W_i^*$ from (4.12) into (4.9) and using Assumption 4.3, the proof follows. □

**Remarks: 1)** A condition under which inequality (4.6) holds is

$$\frac{g_i}{l_i} > 1, \qquad (4.13)$$

for all $i \in I$.

**2)** System (2.2) has a unique type-$n$ equilibrium point. This point is of great importance since it corresponds to nonzero soliton pulse energies in all channels of a WDM transmission system in long distances. □

Up to this point, the equilibrium points of system (2.2) were computed. Next, the stability of these points is studied. The important result to be established in the following is that only the unique type-$n$ equilibrium point of system (2.2) is stable. This is certainly a very desirable situation since it guarantees that the energies of solitons in all channels of a transmission system remain positive in long distance. The realization of this situation is due to guiding filters.

### 4.2. Stability of Equilibrium Points

In studying the stability of equilibrium points of system (2.2), three sets of such points are considered: (i) the origin $\theta_n$ ; (ii) type-$k$ equilibrium points for all $1 \leq k \leq n-1$ ; (iii) the unique type-$n$ equilibrium point.

Using the Jacobian matrix, it is first shown that $\theta_n$ cannot be attained in practice.

**Theorem 4.5:** If $g_i > l_i$ for an $i = 1, 2, \ldots, n$ (respectively, $g_i < l_i$ for all $i = 1, 2, \ldots, n$), then the origin $\theta_n$ is an unstable (a locally asymptotically stable) equilibrium point of system (2.2).

**Proof:** The proof is similar to that of Theorem 3.4, except that now the Jacobian matrix of system (2.2) is computed at $\theta_n$. This matrix turns out to be identical to that in (3.6). Thus, the proof follows. □

**Remark:** Since in practice, for at least one channel the gain rate is larger than the loss rate, by Theorem 4.5, the soliton pulse energies in all channels do not tend to zero. □

Before studying the stability of other equilibrium points of system (2.2), a useful result is established.

**Lemma 4.6:** Consider the matrix

$$A = \begin{bmatrix} -a_1 - b_1 & -a_1 & \cdots & -a_1 \\ -a_2 & -a_2 - b_2 & \cdots & -a_2 \\ \vdots & \vdots & \vdots & \vdots \\ -a_k & -a_k & \cdots & -a_k - b_k \end{bmatrix} \in \mathbb{R}^{k \times k}, \quad (4.14)$$

where $a_i > 0$ and $b_i > 0$ for all $i = 1, 2, \ldots, k$. All eigenvalues of $A$ have negative real parts.

**Proof:** The following positive definite diagonal matrix is chosen:

$$P = diag\left[\frac{1}{a_1}, \frac{1}{a_2}, \ldots, \frac{1}{a_k}\right] \in \mathbb{R}^{k \times k}. \quad (4.15)$$

With this choice, it follows that

$$A^T P + PA = -2Q, \quad (4.16)$$

where

$$Q = \begin{bmatrix} 1 + \frac{b_1}{a_1} & 1 & \cdots & 1 \\ 1 & 1 + \frac{b_2}{a_2} & \cdots & 1 \\ \vdots & \vdots & \vdots & \vdots \\ 1 & 1 & \cdots & 1 + \frac{b_k}{a_k} \end{bmatrix} \in \mathbb{R}^{k \times k}. \quad (4.17)$$

Starting form the $(1, 1)$-element of the matrix $Q$, the principal minors of $Q$ can be computed. The $j$-th principal minor for a $j = 1, 2, \ldots, k$ is

$$\pi_j = \left(\frac{b_1}{a_1}\right)\left(\frac{b_2}{a_2}\right)\cdots\left(\frac{b_j}{a_j}\right)\left(\frac{a_1}{b_1} + \frac{a_2}{b_2} + \ldots + \frac{a_j}{b_j} + 1\right). \quad (4.18)$$

Since $\pi_j > 0$ for all $j = 1, 2, \ldots, k$, the matrix $Q$ is positive definite. Therefore, by Theorem

42 in Ref. [10, p. 199] or Theorem 1.2 in Ref. [12, p. 6], all eigenvalues of $A$ have negative real parts. □

The stability of type-$k$ equilibrium points for all $1 \leq k \leq n-1$ is now studied.

**Theorem 4.7:** For any integer $1 \leq k \leq n-1$, all type-$k$ equilibrium points of system (2.2) are unstable.

**Proof:** For a $1 \leq k \leq n-1$, without loss of generality, let system (2.2) have a type-$k$ equilibrium point whose first $k$ coordinates are positive. That is, let the equilibrium point be

$$W^* = (W_1^*, W_2^*, \ldots, W_k^*, 0, \ldots, 0) \in \mathbb{R}_+^n. \tag{4.19}$$

The Jacobian matrix of system (2.2) at $W^*$ is obtained as:

$$J(W^*) = \begin{bmatrix} M_{11} & M_{12} \\ \Theta & M_{22} \end{bmatrix} \in \mathbb{R}^{n \times n}, \tag{4.20}$$

where $\Theta$ is the $(n-k) \times k$ zero matrix and

$$M_{11} = \begin{bmatrix} -\dfrac{Cg_1 W_1^*}{\Delta^2} - \dfrac{s_1 W_1^*}{D_1} & -\dfrac{Cg_1 W_1^*}{\Delta^2} & \cdots & -\dfrac{Cg_1 W_1^*}{\Delta^2} \\ -\dfrac{Cg_2 W_2^*}{\Delta^2} & -\dfrac{Cg_2 W_2^*}{\Delta^2} - \dfrac{s_2 W_2^*}{D_2} & \cdots & -\dfrac{Cg_2 W_2^*}{\Delta^2} \\ \vdots & \vdots & \vdots & \vdots \\ -\dfrac{Cg_k W_k^*}{\Delta^2} & -\dfrac{Cg_k W_k^*}{\Delta^2} & \cdots & -\dfrac{Cg_k W_k^*}{\Delta^2} - \dfrac{s_k W_k^*}{D_k} \end{bmatrix} \in \mathbb{R}^{k \times k}, \tag{4.21a}$$

$$M_{12} = \begin{bmatrix} -\dfrac{Cg_1 W_1^*}{\Delta^2} & -\dfrac{Cg_1 W_1^*}{\Delta^2} & \cdots & -\dfrac{Cg_1 W_1^*}{\Delta^2} \\ -\dfrac{Cg_2 W_2^*}{\Delta^2} & -\dfrac{Cg_2 W_2^*}{\Delta^2} & \cdots & -\dfrac{Cg_2 W_2^*}{\Delta^2} \\ \vdots & \vdots & \vdots & \vdots \\ -\dfrac{Cg_k W_k^*}{\Delta^2} & -\dfrac{Cg_k W_k^*}{\Delta^2} & \cdots & -\dfrac{Cg_k W_k^*}{\Delta^2} \end{bmatrix} \in \mathbb{R}^{k \times (n-k)}, \tag{4.21b}$$

$$M_{22} = \mathrm{diag}\left[\dfrac{g_{k+1}}{\Delta} - l_{k+1},\ \dfrac{g_{k+2}}{\Delta} - l_{k+2},\ \ldots,\ \dfrac{g_n}{\Delta} - l_n\right] \in \mathbb{R}^{(n-k) \times (n-k)}, \tag{4.21c}$$

and $\Delta$ is that in (4.5) for $I = \{1, 2, \ldots, k\}$.

Clearly, the eigenvalues of $J(W^*)$ are those of $M_{11}$ and $M_{22}$. By Assumption 4.3, all eigenvalues of $M_{22}$ are positive. (Note that by Lemma 4.6, all eigenvalues of $M_{11}$ have negative real parts, although, this information is not necessary for the proof.) Therefore, $W^*$ is an unstable equilibrium point of system (2.2). □

Finally, the stability of the unique type-$n$ equilibrium point of system (2.2) is established.

**Theorem 4.8:** The unique type-$n$ equilibrium point of system (2.2) is locally asymptotically stable.

**Proof:** The coordinates of the unique type-$n$ equilibrium point of system (2.2) are all positive. That is, the equilibrium point has the following form:

$$W^* = (W_1^*, W_2^*, \ldots, W_n^*) \in \mathbb{R}_+^n. \tag{4.22}$$

The Jacobian matrix of system (2.2) at $W^*$ is obtained as:

$$J(W^*) = \begin{bmatrix} -\dfrac{Cg_1 W_1^*}{\Delta^2} - \dfrac{s_1 W_1^*}{D_1} & -\dfrac{Cg_1 W_1^*}{\Delta^2} & \cdots & -\dfrac{Cg_1 W_1^*}{\Delta^2} \\ -\dfrac{Cg_2 W_2^*}{\Delta^2} & -\dfrac{Cg_2 W_2^*}{\Delta^2} - \dfrac{s_2 W_2^*}{D_2} & \cdots & -\dfrac{Cg_2 W_2^*}{\Delta^2} \\ \vdots & \vdots & \vdots & \vdots \\ -\dfrac{Cg_n W_n^*}{\Delta^2} & -\dfrac{Cg_n W_n^*}{\Delta^2} & \cdots & -\dfrac{Cg_n W_n^*}{\Delta^2} - \dfrac{s_n W_n^*}{D_n} \end{bmatrix} \in \mathbb{R}^{n \times n}. \tag{4.23}$$

By Lemma 4.6, all eigenvalues of $J(W^*)$ have negative real parts. Hence, $W^*$ is locally asymptotically stable. □

Theorem 4.8 shows the success of guiding filters in keeping the energies of solitons positive in all channels of a transmission system in long distances. Nevertheless, this theorem presents a local stability result. Attempts to prove the global stability of the unique type-$n$ equilibrium point of system (2.2) have been unsuccessful. However, in the following, it is proved that the solution of system (2.2) remains bounded.

**Theorem 4.9:** The solution of system (2.2) is bounded. More precisely, trajectories corresponding to the solutions while traverse in the positive orthant $\mathbb{R}_+^n$ remain inside or on the ellipsoid given by

$$\frac{W_1^2}{g_1 D_1/s_1} + \frac{W_2^2}{g_2 D_2/s_2} + \ldots + \frac{W_n^2}{g_n D_n/s_n} = \frac{1}{C} . \tag{4.24}$$

**Proof:** By Theorem 2.1, trajectories corresponding to the solutions of system (2.2) are in the positive orthant. However, they remain bounded as it is proved in the following. Consider the function $E : \mathbb{R}^n \rightarrow \mathbb{R}_+$ given by

$$E(z) = W_1(z)/g_1 + W_2(z)/g_2 + \ldots + W_n(z)/g_n , \tag{4.25}$$

for all $z \in [0, z_f = \infty)$. By Theorem 2.1, $E(\cdot)$ is a non-negative function. The derivative of $E(\cdot)$ along the solution of system (2.2) is

$$\begin{aligned}
\frac{dE}{dz} &= \frac{W_1 + W_2 + \ldots + W_n}{1 + C(W_1 + W_2 + \ldots + W_n)} \\
&\quad - \left( \frac{l_1}{g_1} W_1 + \frac{l_2}{g_2} W_2 + \ldots + \frac{l_n}{g_n} W_n \right) \\
&\quad - \left( \frac{W_1^2}{g_1 D_1/s_1} + \frac{W_2^2}{g_2 D_2/s_2} + \ldots + \frac{W_n^2}{g_n D_n/s_n} \right).
\end{aligned} \tag{4.26}$$

By Theorem 2.1: (i) $\sum_{i=1}^{n} W_i \geq 0$, and hence the first term on the right-hand side of (4.26) is less than or equal to $1/C$; (ii) the second term of (4.26) is non-positive. Thus,

$$\frac{dE}{dz} \leq \frac{1}{C} - \left( \frac{W_1^2}{g_1 D_1/s_1} + \frac{W_2^2}{g_2 D_2/s_2} + \ldots + \frac{W_n^2}{g_n D_n/s_n} \right). \tag{4.27}$$

Clearly, $dE/dz < 0$ for all points outside the ellipsoid in (4.24). Thus, trajectories corresponding to the solutions of system (2.2) remain inside or on the ellipsoid. $\square$

## 5. Conclusions

In this paper, the regulating effect of guiding filters on the energies of solitons in multi-channel wave-length division multiplexing (WDM) transmission systems was rigorously studied. This study was based on two mathematical models that represent the evolution of energies of solitons in channels of transmission systems in the absence and presence of guiding filters. The models are nonlinear dynamical systems whose states are the energies of solitons. Using techniques from the theory of dynamical systems, it was first shown that in the absence of guiding filters the energies of solitons cannot remain nonzero in long distances, and all, except one of them, tend to zero. In the presence of guiding filters, however, it was proved that in all transmission channels the energies of solitons remain positive in long distances. The results of this paper are rigorous mathematical justifications of numerical and experimental results available in the literature.

# References

[1] L. F. Mollenauer, J. P. Gordon, and S. G. Evangelides, The sliding-frequency guiding filter: an improved form of soliton jitter control, *Optics Letters*, 17 (22) (1992), 1575-1577.

[2] L. F. Mollenauer, P. V. Mamyshev, and M. J. Neubelt, Measurement of timing jitter in filter-guided soliton transmission at 10 Gbits/s and achievement of 375 Gbits/s-Mm, error free, at 12.5 and 15 Gbits/s, *Optics Letters*, 19 (10) (1994), 704-706.

[3] P. V. Mamyshev and L. F. Mollenauer, Stability of soliton propagation with sliding-frequency guiding filters, *Optics Letters*, 19 (24) (1994), 2083-2085.

[4] A. Hasegawa and Y, Kodama, *Solitons in Optical Communications*, Oxford University Press, Oxford, UK, 1995.

[5] P. V. Mamyshev and L. F. Mollenauer, Wavelength-division-multiplexing channel energy self-equalization in a soliton transmission line by guiding filters, *Optics Letters*, 21 (20) (1996), 1658-1660.

[6] L. F. Mollenauer, P. V. Mamyshev, and M. J. Neubelt, Demonstration of soliton WDM transmission at 6 and 7 X 10 Gbit/s, error free over transoceanic distances, *Electronics Letters*, 32 (15) (1996), 471-473.

[7] L. F. Mollenauer, P. V. Mamyshev, and T. A. Strasser, Guiding filters for massive wavelength division multiplexing in soliton transmission, *Optics Letters*, 22 (1) (1997), 1621-1623.

[8] L. F. Mollenauer and J. P. Gordon, *Solitons in Optical Fibers: Fundamentals and Applications*, Elsevier, Boston, MA, 2006.

[9] V. V. Chepyzhov and M. I. Vishik, *Attractors for Equations of Mathematical Physics*, American Mathematical Society, Providence, RI, 2002.

[10] M. Vidyasagar, *Nonlinear Systems Analysis*, Prentice Hall, Englewood Cliffs, NJ, 1993, 2nd edition.

[11] S.-B. Hsu, *Ordinary Differential Equations with Applications*, World Scientific, Singapore, 2006.

[12] Z. Gajic and M. T. J. Qureshi, *Lyapunov Matrix Equation in System Stability and Control*, Academic Press, San Diego, CA, 1995.